\documentclass[prx,twocolumn,english,showpacs,nofootinbib]{revtex4-1}
\usepackage{amssymb}
\usepackage{color}
\usepackage{graphicx}
\usepackage{bbold}
\usepackage{bm}
\usepackage{amsmath}
\usepackage{amsfonts}
\usepackage{amssymb}
\usepackage{braket}
\usepackage{units}
\usepackage{csquotes}
\usepackage{upgreek}
\usepackage{xcolor}
\usepackage{fancyhdr}
\begin{document}

\title{Coherent Manipulation of Spin Correlations in the Hubbard Model}

\author{N. Wurz,$^1$ C. F. Chan,$^1$  M. Gall,$^1$ J. H. Drewes,$^1$ E. Cocchi,$^{1,2}$ L. A. Miller,$^{1,2}$\\   D. Pertot,$^1$ F. Brennecke,$^1$ and M. K\"ohl$^1$}

\affiliation{$^1$Physikalisches Institut, University of Bonn, Wegelerstra{\ss}e 8, 53115 Bonn, Germany\\
$^2$\mbox{Cavendish Laboratory, University of Cambridge, JJ Thomson Avenue, Cambridge CB3 0HE, United Kingdom}}

\begin{abstract}
We coherently manipulate spin correlations in a two-component atomic Fermi gas loaded into an optical lattice using spatially and time-resolved Ramsey spectroscopy combined with high-resolution \textit{in situ} imaging. This novel technique allows us not only to imprint spin patterns but also to probe the static magnetic structure factor at arbitrary wave vector, in particular the staggered structure factor. From a measurement along the diagonal of the $1^\mathrm{st}$ Brillouin zone of the optical lattice, we determine the magnetic correlation length and the individual spatial spin correlators. At half filling, the staggered magnetic structure factor serves as a sensitive thermometer for the spin temperature, which we employ to study the thermalization of spin and density degrees of freedom during a slow quench of the lattice depth.
\end{abstract}
\maketitle

Quantum magnetism arises in spin-1/2 many-body systems from the exchange interaction and has been studied since the early days of quantum physics \cite{Auerbach1994}. The discovery of high-temperature superconductivity in the cuprates \cite{Dagotto1994,Lee2006} has triggered renewed interest in quantum magnetism owing to a subtle interplay between doping, dimensionality and lattice geometry. The two-dimensional (2D) Hubbard model is the most elementary, yet analytically intractable model that is believed to capture essential aspects of high-temperature superconductivity. Ultracold fermionic atoms loaded into the periodic potential of an optical lattice constitute a highly tunable and faithful experimental realization of the Hubbard model and hold the promise to provide deeper insight into its low-temperature phases \cite{Esslinger2010}.

{The magnetic properties of a many-body system are characterized by the (static) magnetic structure factor $S(\bm{q})$, which quantifies the Fourier spectrum of spatial magnetic correlations. 	The emergence of magnetic order is signaled by a peak in the spin structure factor at a certain wave vector, e.g. at $\bm{q}_\mathrm{AFM} = \left(\pi/a, \pi/a\right)$ for a 2D antiferromagnet (AFM) on a square lattice with lattice constant $a$.} In solid-state systems, the structure factor can be inferred from neutron or X-ray scattering experiments \cite{Squires1980,Platzman1970}. A related technique was demonstrated in a three-dimensional optical lattice experiment using Bragg scattering of laser light, providing access to the magnetic structure factor at a fixed wave vector and averaged over regions of different filling \cite{Hart2015}. More recently, ultracold quantum gas experiments have observed antiferromagnetic correlations in the 2D Hubbard model by measuring site-resolved spin correlations \cite{Parsons2016,Boll2016,Cheuk2016b,Brown2016}, from which the magnetic structure factor at half filling was computed \cite{Mazurenko2016}. Additionally, the uniform magnetic structure factor was determined for arbitrary filling by detecting fluctuations of the magnetization \cite{Drewes2017}.

\begin{figure}[hb!]
	\includegraphics[width = 0.90\columnwidth,clip=true]{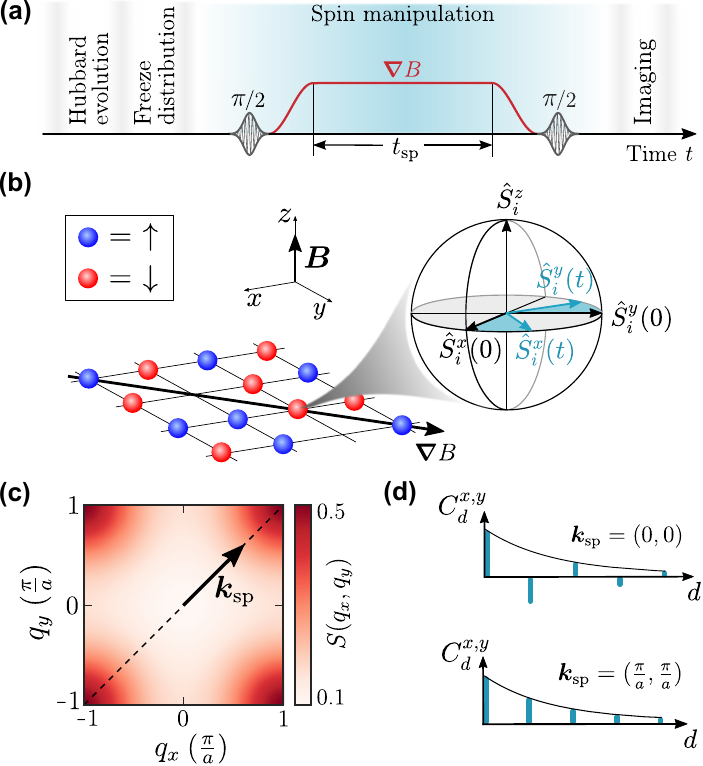}
	\caption{(a) Experimental scheme to imprint spatially periodic spin patterns onto fermionic spin-1/2 atoms loaded into an optical  lattice potential realizing the 2D Hubbard Hamiltonian. After freezing the atomic motion, a Ramsey-type sequence is applied, which is composed of two $\pi/2$ RF pulses enclosing the evolution in a magnetic field gradient $\bm{\nabla}B$ for a variable time $t_\mathrm{sp}$.  		(b) The time evolution of the spin operator $\hat{\bm{S}}_i$ can be considered as a precession around the magnetic field direction with a locally varying Larmor frequency.
		(c) Spin structure factor $S(\bm{q})$ at half filling plotted as a function of $\bm{q}$ in the $1^\mathrm{st}$ BZ assuming an exponential decay of the spatial correlators with correlation length $\xi = 0.45\, a$.  (d) {Transverse spin correlators as a function of distance $d$ for the unperturbed and spin-manipulated two-dimensional antiferromagnet.}}
	\label{fig1}
\end{figure}

Here, we present the coherent manipulation of spin correlations as a novel probe for the magnetic structure factor $S(\bm{q})$ of a spin-1/2 Fermi gas in an optical lattice at arbitrary wave vector $\bm{q}$. The key idea is to expose the atomic cloud to a magnetic field gradient in order to coherently manipulate its spin correlations (see Figs.~1(a) and 1(b)). In the magnetic field gradient, the spins  precess in the transverse $xy$-plane at a rate that depends on the local magnetic field strength. Therefore, the time evolution creates a periodic pattern of the transverse magnetization with well-defined wave vector $\bm{k}_\mathrm{sp}$ \cite{Koschorreck2013,bardon2014transverse,hild2014far,luciuk2017observation}. Subsequently, we {rotate} the transverse magnetization onto the  longitudinal $z$-magnetization by applying a $\pi/2$ radiofrequency (RF) pulse. Now, by performing spin-selective \textit{in situ} imaging in the $z$-basis, we measure the \textit{uniform} magnetic structure factor \cite{Drewes2017} of the spin-manipulated system and hence our experimental protocol detects the magnetic structure factor of the initially prepared gas at wave vector $\bm{q} = \bm{k}_\mathrm{sp}$. Note that this scheme does not require a site-resolving imaging technique even for measuring correlations at the reciprocal lattice vector.

In this work, we use this technique to record the magnetic structure factor of the 2D Hubbard model along the diagonal of the $1^\mathrm{st}$ Brillouin zone (BZ) in a system of ultracold $^{40}\mathrm{K}$ atoms loaded into an optical lattice potential. 
In a single-band approximation, the Hubbard Hamiltonian in the grand canonical ensemble is given by
\begin{equation}
\hat{H} = -t\sum_{\langle i,j\rangle,\sigma}\hat{c}_{i,\sigma}^\dag\hat{c}_{j,\sigma}+U\sum_i \hat{n}_{i,\uparrow} \hat{n}_{i,\downarrow} - \mu\sum_{i,\sigma}\hat{n}_{i,\sigma}
\end{equation}
where $t$ denotes the tunneling amplitude of fermionic particles with spin $\sigma = \uparrow, \downarrow$ between neighboring lattice sites $i$ and $j$, and $U$ denotes the interaction strength between fermions of opposite spin residing on the same lattice site. The lattice filling $n = (\langle \hat{n}_\uparrow\rangle+\langle \hat{n}_\downarrow\rangle)/2$ is controlled via the chemical potential $\mu$, which spatially varies in the experiment owing to the external confinement \cite{Cocchi2016}. For half filling  $n = 1/2$, corresponding to one particle per site on average, and strong interactions $U \gg t$, the Hubbard model realizes a Mott insulating phase with antiferromagnetic correlations for temperatures below the superexchange interaction strength $J = 4 t^2/U$. In two spatial dimensions, a long-range-ordered AFM phase is realized only in the zero-temperature limit \cite{Mermin1966}.

\begin{figure*}
 \includegraphics[width=0.9\textwidth,clip=true]{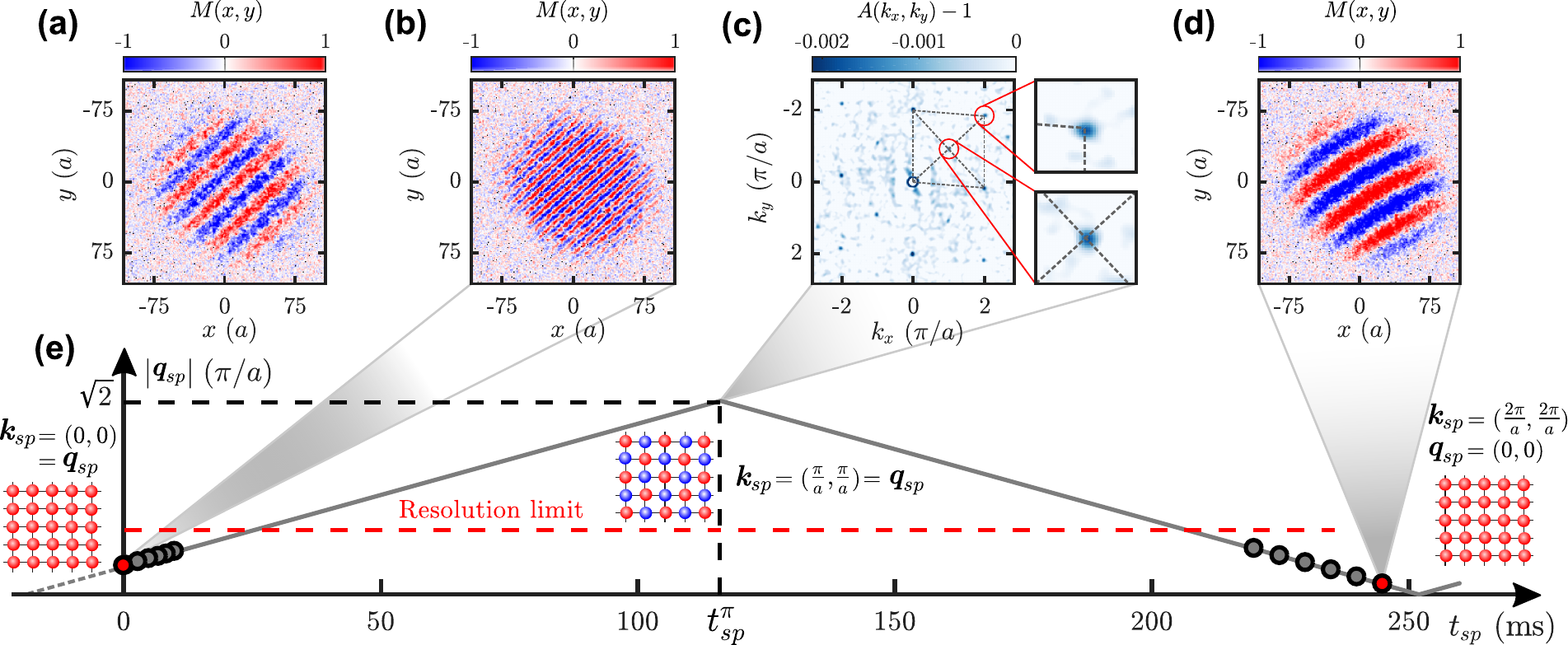}
 \caption{Imprinting spin patterns onto a spin-polarized cloud. The magnetic field gradient is ramped up to $|\bm{\nabla} B| = \unit[0.64]{G/cm}$, except for (a), where $|\bm{\nabla} B| = \unit[0.1]{G/cm}$. (a,b) Contrast $M(x,y)$ for spin patterns observed  at different gradient strengths, and $t_\mathrm{sp} = \unit[0.1]{ms}$. (c) Normalized density-density correlation signal  (average of 200 realizations) of a single spin component after trap release and ballistic expansion for $\unit[4]{ms}$. The additional correlation dips around $(\pm\, {\pi}/{a},\pm \, \pi/a)$ demonstrate the realization of a checkerboard spin pattern with  wave vector $\bm{q}_{\mathrm{AFM}}$. (d) Observed spin pattern with small wave vector after Bragg reflection. The contrast is still high, suggesting that single-particle coherence is maintained. A small residual magnetic field curvature of around $\unit[11]{nG/\upmu m^2}$ is deduced from a fit to the fringe pattern. 
 	(e) We plot the magnitude of the extracted wave vector $|\bm{q}_{\mathrm{sp}}|$ versus the duration $t_\mathrm{sp}$ of the spin evolution in the magnetic field gradient. The dashed red line shows the resolution limit of our imaging setup. A combined fit (solid line)  provides a precise calibration to imprint spin patterns with arbitrary wave vectors along the diagonal of the $1^\mathrm{st}$ BZ and to extract the turning point $t_\mathrm{sp}^\pi$.}
\label{fig2}
\end{figure*}
The magnetic properties of the Hubbard model are characterized by the spin (or magnetic) structure factor \begin{equation}\label{eq:struct_fact}
	S(\bm{q}) = \frac{1}{N}\sum_{i,j} e^{-i \bm{q}\cdot \bm{r}_{ij}}C^z_{ij}\, .
\end{equation} Here, $\bm{r}_{ij}=\bm{r}_{j}-\bm{r}_{i}$ is the vector connecting the lattice sites $i$ and $j$, $N$ is the number of lattice sites, and $C^\nu_{ij} = \braket{ \hat{S}_i^\nu \hat{S}_j^\nu}-\braket{ \hat{S}_i^\nu} \braket{\hat{S}_j^\nu}$ denotes the spin correlator between sites $i$ and $j$ along the spin direction $\nu=x,y,z$. The local spin-1/2 operator $\hat{\bm{S}}_j$ acting on site $j$ is defined by $\hat{S}^x_j = (\hat{c}_{j,\uparrow}^\dag \hat{c}_{j,\downarrow} + \hat{c}_{j,\downarrow}^\dag \hat{c}_{j,\uparrow})/2$, $\hat{S}^y_j = (\hat{c}_{j,\uparrow}^\dag \hat{c}_{j,\downarrow} - \hat{c}_{j,\downarrow}^\dag \hat{c}_{j,\uparrow})/(2i)$, and $\hat{S}^z_j = (\hat{n}_{j,\uparrow}-\hat{n}_{j,\downarrow})/2$. 
 Note that for an SU(2) symmetric system the second term of the correlator vanishes, and the spin correlations are isotropic {$C^\nu_{ij}\equiv C_{ij}$}. Owing to the alternating sign of AFM correlations, $C_{ij} \propto e^{i \bm{q}_\mathrm{AFM} \cdot \bm{r}_{ij}}$,  the spin structure factor at half filling exhibits a maximum at the corners of the first BZ, i.e.~for $\bm{q}_\mathrm{AFM} = (\pi/a,\pi/a)$, where $a$ denotes the lattice constant, and reaches a minimum at $\bm{q} = 0$ (Fig.~1(c)). Whereas the \emph{staggered} spin structure factor $S(\bm{q}_\mathrm{AFM})$ rises exponentially with decreasing temperature, the \emph{uniform} structure factor $S(\bm{q}=0)$ diminishes in the zero-temperature limit \cite{hirsch1985two}.

In our experiments we prepare an ultracold cloud of fermionic $^{40}\mathrm{K}$ atoms in a balanced mixture of the two lowest hyperfine states $\ket{\uparrow} = \ket{F=9/2, m_F=-7/2}$ and $\ket{\downarrow} = \ket{F=9/2, m_F=-9/2}$ \cite{Cocchi2016}. The atoms are loaded into the lowest band of an anisotropic three-dimensional optical lattice potential with strongly suppressed tunneling along the vertical direction. Within the horizontal planes an almost square lattice potential is applied with a lattice constant of $a=\unit[532]{nm}$ and a depth of $6\, {E_\mathrm{rec}}$ resulting in a tunneling amplitude of $t/h = \unit[224(6)]{Hz}$. Here, $E_\mathrm{rec} = \frac{h^2}{8 m a^2}$ denotes the recoil energy with atomic mass $m$ and Planck constant $h$. The interaction strength $U$ is set to $U = 8.2(5)\, t$ by tuning a vertically aligned magnetic bias field in the vicinity of the Feshbach resonance located at $\unit[202]{G}$. After preparing the atoms in the lattice, we freeze their positions by quickly (within \unit[1]{ms}) ramping the horizontal lattice depth to a value of $60\, {E_\mathrm{rec}}$. We remove atoms residing on doubly occupied lattice sites as they do not contribute to the magnetic properties of the system \cite{Drewes2017}. 
In order to spin-selectively record the in-trap density distributions in a single horizontal layer we perform RF tomography in a vertical magnetic field gradient. Finally, after manipulating the many-body spin state using our novel scheme, we successively take absorption images of both spin components with a spatial resolution of $\unit[1.2]{\upmu m}$ \cite{Drewes2017}.

The experimental sequence for writing periodic spin patterns into the frozen lattice system as well as manipulating its spin correlations is shown in Fig.~1(a). In between two RF pulses, which induce $\pi/2$ rotations of the $\ket{\uparrow}$ and $\ket{\downarrow}$  states, an in-plane magnetic field gradient $\bm{\nabla} B$ is applied. In the Heisenberg picture this induces a precession of the transverse spin components $\hat{S}^x_i$ and $\hat{S}^y_i$ with a spatially varying Larmor frequency $\omega_{i}$. Consequently, two spins at lattice sites $i$ and $j$ acquire a phase difference  $\phi_{ij} = \int \left( \omega_{j} - \omega_i\right) dt = \gamma\int \bm{\nabla} B(t) \cdot \bm{r}_{ij}\, dt$ in their transverse spin orientation. Here, $\gamma$ denotes the differential gyromagnetic ratio between the $\ket{\uparrow}$ and $\ket{\downarrow}$ states. By adjusting the magnetic field gradient $\bm{\nabla} B$ and the spin evolution time, we imprint phase patterns in spin space with arbitrary wave vector $\bm{k}_\mathrm{sp} = \gamma\int   \bm{\nabla} B (t) \, dt $. While this leaves the vertical spin correlators  $C^z_{ij}$ unchanged, the transverse spin correlators $C^{x,y}_{ij}$ of the spin-manipulated system evolve in the magnetic field gradient according to
\begin{equation}\label{eq:oscillating_corr}
	C^{x,y}_{ij}(\bm{k}_\mathrm{sp}) = \cos(\bm{k}_\mathrm{sp} \cdot \bm{r}_{ij})C_{ij}\, .
\end{equation} This is a direct consequence of the local precession of the transverse spin components and the SU(2) symmetry of the Hubbard model, which makes cross-correlators of the form $\langle \hat{S}^\nu_i\hat{S}^\mu_j\rangle$ vanish for all $\mu\neq \nu$. In particular, when aligning the gradient direction with a lattice diagonal, the transverse correlators, starting at $\bm{k}_\mathrm{{sp}} = (0,0)$ from a state with short-ranged antiferromagnetic correlations, successively unwind their staggered pattern and take equal signs for $\bm{k}_\mathrm{sp} = \bm{q}_\mathrm{AFM}$, as shown in Fig.~1(d).
 
In order to access the time-evolved transverse correlators, we apply a $\pi/2$-pulse, rotating a transverse spin direction into the vertical spin direction, which serves as the measurement basis of the spin-selective imaging. The spatially integrated spin correlations, as extracted from images of the density distributions of both spin components $n_{\uparrow,\downarrow}(\bm{r})$, reveal the spin structure factor of the initially prepared gas at wave vector $\bm{q} = \bm{k}_\mathrm{sp}$: $  \frac{1}{N}\sum_{i,j}C^{x,y}_{ij}(\bm{k}_\mathrm{sp}) = S(\bm{k}_\mathrm{sp})$. Here, we made use of the invariance $C^\nu_{ij}=C^\nu_{ji}$, which according to Eq.~\ref{eq:oscillating_corr} remains intact even when a spin pattern of arbitrary wave vector is imprinted. Since the wave vector $\bm{k}_\mathrm{sp}$ is tunable by adjusting the magnetic field gradient and the spin evolution time $t_\mathrm{sp}$, this scheme provides a powerful tool  to probe the spin structure factor $S(\bm{q})$ at arbitrary  $\bm{q}$. 

For the calibration of our technique, we start with a spin-polarized cloud of atoms in $\ket{\downarrow}$ and imprint a spin pattern with wave vector $\bm{k}_\mathrm{sp}$. Typical images of the observed contrast $M(\bm{r}) = \frac{ n_\uparrow(\bm{r})-n_\downarrow(\bm{r})}{ n_\uparrow(\bm{r})+n_\downarrow(\bm{r})}$ showing spin patterns with different wave vectors are displayed in Figs.~2(a) and 2(b).  Owing to the discrete nature of the lattice, the  wave vector $\bm{q}_\mathrm{sp}$ extracted from the Fourier transformation of $M(\bm{r})$   is bounded to the $1^\mathrm{st}$ BZ. Our optical resolution allows us to resolve spin patterns with wave vectors up to $|\bm{q}_\mathrm{sp}|\approx 2\pi/(5a)$. For evolution times $t_\mathrm{sp}>\unit[220]{ms}$, we again observe spin patterns with decreasing wave vector owing to Bragg reflection at the boundary of the $1^\mathrm{st}$ BZ (see Figs.~2(d) and 2(e)). When the wave vector is aligned with the lattice diagonal as intended, we observe a recurrence of the spin polarized gas at $\bm{k}_\mathrm{sp} = (\frac{2\pi}{a},\frac{2\pi}{a})$, which we use for fine adjustment of the gradient direction. From a combined fit to the observed evolution of the wave vector of the spin pattern, we extract the evolution time $t_\mathrm{sp}^\pi$ required for realizing a checkerboard spin pattern with $\bm{k}_\mathrm{sp} = \bm{q}_\mathrm{AFM}$ as illustrated in Fig.~2(e). The finite wave vector at $t_\mathrm{sp}=0$ is caused by the spin evolution during the time intervals when the magnetic field gradient is turned on and off (see Fig.~1(a)).

In addition, we confirmed the existence of a spin pattern with wave vector $\bm{q}_\mathrm{sp} = \bm{q}_\mathrm{AFM}$ by recording the density-density correlations (\enquote{noise correlations}) $A(\bm{k})\propto \int d^2\bm{q}\langle \hat{n}_\downarrow (\bm{q}) \hat{n}_\downarrow (\bm{q}+\bm{k})\rangle$ of one spin component after ballistic expansion of the cloud in the 2D plane \cite{altman2004probing,Foelling2005,Rom2006}. Imprinting a checkerboard spin pattern effectively rotates and shrinks the area of the reciprocal unit cell by a factor of ${2}$. Therefore, we observe additional anti-bunching dips at $\left(\pm \pi/a, \pm \pi/a \right)$ as compared to the case of a spin-polarized 2D plane (see Fig.~2(c)).
\begin{figure}
 \includegraphics[width=0.9\columnwidth,clip=true]{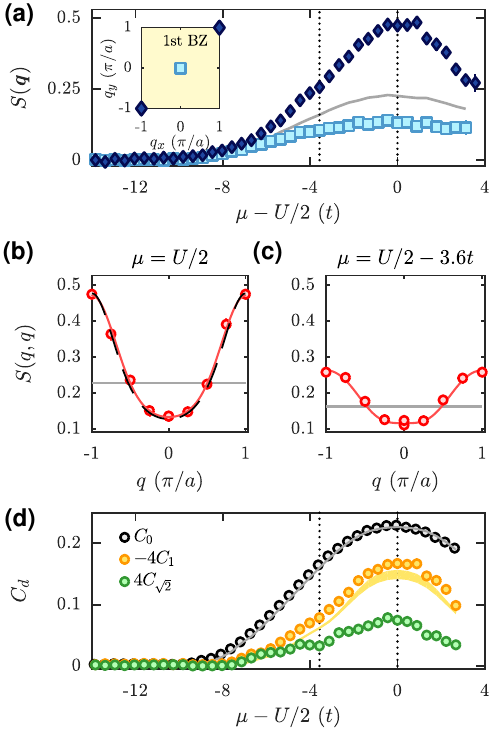}
 \caption{(a) Uniform (squares) and staggered (diamonds) spin structure factor recorded as a function of chemical potential $\mu$. The solid gray line shows the measured local moment $C_0$. The two dashed lines highlight the potential bins used for Figs.~3(b) and 3(c). (b,c) Spin structure factor (circles) and local moment (solid gray line) recorded along the diagonal of the BZ at half filling $\mu=U/2$ and away from half filling $\mu = U/2-3.6\,t$, respectively. {We plot the fit using the Fourier series given in Eq.~\ref{eq:fourier} (solid red line), and additionally, at half filling, the fit assuming an exponential decay of the magnitude of the spatial correlators according to Eq.~\ref{eq:expo} (black dashed line).} (d) Extracted spin correlators $C_d$ (circles) as a function of chemical potential. The gray and yellow shadings show NLCE data for a temperature interval $k_B T = [0.54\, ,0.6]\,t$ of the local moment and nearest-neighbor correlator, respectively.}
\label{fig3}
\end{figure}

We now present a measurement of the spin structure factor $S(\bm{q})$ in the 2D Hubbard model resulting from the coherent evolution of spin correlations. Fig.~3(a) shows experimental data of the uniform spin structure factor \mbox{$S(\bm{q}=0)$} (recorded without manipulating the spins) and the staggered spin structure factor $S(\bm{q}_\mathrm{AFM})$ as a function of the chemical potential $\mu$, which is inferred in a local density  approximation (LDA) from the precisely calibrated trapping potential \cite{Cocchi2016}. As a reference, we also show the measured local moment $C_{ii} \equiv C_{0} = (\langle \hat{s}_\uparrow\rangle+\langle \hat{s}_\downarrow\rangle)/4$ in gray, which reaches a maximum at half filling $\mu = U/2$ and is extracted from the averaged densities of singly occupied lattice sites (\enquote{singles}) $\langle \hat{s}_{\sigma}\rangle = \langle \hat{n}_{\sigma}-\hat{n}_{\uparrow}\hat{n}_{\downarrow}\rangle$ for $\sigma = \uparrow, \downarrow$ \cite{Drewes2017}. We note, that we have confirmed the SU(2) symmetry of our experimental implementation of the 2D Hubbard model by removing the first $\pi/2$-pulse of the Ramsey-type sequence (see Fig.~1(a)), which yielded the same values for the staggered and uniform structure factors. In previous work we had concentrated on measuring the uniform spin structure factor (squares in Fig.~3(a)), which is smaller than the local moment owing to the presence of negative nearest-neighbor correlations \cite{Drewes2017}. In contrast, the staggered spin structure factor  (diamonds in Fig.~3(a)) is observed to exceed the local moment by more than the mismatch between the local moment and the uniform structure factor, since all spatial correlators add up constructively. This asymmetry with respect to the local moment clearly indicates the presence of beyond nearest-neighbor AFM spin correlations.
Since the staggered spin structure factor is very sensitive to changes in temperature, we employ it as a local thermometer. By direct comparison to calculations using {numerical linked-cluster expansion (NLCE)} for the staggered structure factor  {at half filling and $U/t = 8$} \cite{Khatami2011}, we deduce a temperature of $k_B T_s=0.57(3)t$ in the spin sector at half filling. The global density temperature $k_B T_d = 0.63(3)t$, obtained from fitting NLCE data to the singles density profiles, yields a similar result.

In Figs.~3(b) and 3(c), we show the spin structure factor $S(\bm{q})$ measured along the diagonal of the $1^\mathrm{st}$ BZ for different fillings. At half filling, the structure factor exhibits a minimum at $\bm{q} = 0$ and peaks at $\bm{q}_\mathrm{AFM}$, as expected. We observe a qualitatively similar $q$-dependence of the structure factor away from half filling, however, the buildup of spin correlations is suppressed. At half filling, the Hubbard model with strong repulsive interactions can be mapped onto the Heisenberg model \cite{Auerbach1994}, in which the magnitude of spin correlations decays exponentially with a characteristic spin correlation length $\xi$.
Additionally, in a homogeneous system, the spatial spin correlators $C_{ij}$ of the unperturbed spin state depend on the distance $d = |\bm{r}_{ij}|/a$ only. Therefore, at half filling we obtain 
\begin{equation}\label{eq:expo}
	|C_{{i}{j}}| \equiv |C_d| \propto e^{-{d a}/{\xi}}\, ,
\end{equation} %(-1)^{d_x+d_y}
which we use to model $S(q,q)$ according to Eq.~\ref{eq:struct_fact}. We fit this model to the measured structure factor (black dashed line in Fig.~3(b)) and extract a correlation length of $\xi = 0.43(3)\, a$ at half filling.

Recording the spin structure factor as a function \mbox{of $\bm{q}$} further provides access  to the individual spatial spin correlators $C_d$. The value of the correlation length deduced at half filling suggests that spin correlators with $d\geq \sqrt{8}$ do not contribute significantly to the measured spin structure factor. Rewriting $S(q,q)$ along the diagonal of the BZ as a Fourier series we obtain 
\begin{equation}\label{eq:fourier}
	S(q,q) \approx \sum_{n=0}^3 f_n \cos(n q a)
\end{equation} with $f_0 = C_{0}+2 C_{\sqrt{2}}$, $f_1 = 4C_{1}+4C_{\sqrt{5}}$, $f_2 = 2 C_{\sqrt{2}}+4 C_{2}$, and $f_3 = 4 C_{\sqrt{5}}$. Using the independent measurement of the local moment $C_0$ from the density profile alone we are able to extract all higher spin correlators $C_d$ up to a distance of $d = \sqrt{5}$ from the first {four} Fourier components of the measured structure factor. The first three spin correlators are shown in Fig.~3(d) as a function of chemical potential.
We compare the local moment $C_0$ and the nearest-neighbor correlator $C_1$ to data from NLCE calculations for a temperature interval of $k_B T_s = [0.54,0.6] t$. The next-to-nearest neighbor correlator $C_{\sqrt{2}}$ is observed to contribute significantly to the measured spin structure factor and possesses the correct positive sign. The shoulder that is visible  at low filling in the outer region of the  trapping potential is attributed to limitations of LDA, since the atoms experience a larger spatial inhomogeneity owing to the increasing steepness of the trapping potential. This is the more prominent the more long-ranged the quantities become that we extract. The obtained values of $C_2$ and $C_{\sqrt{5}}$ are not shown, because they are mostly consistent with zero within the $1\sigma$ error.

Finally, we employ the staggered structure factor at half filling as a sensitive and local thermometer to study the thermalization of spin and density degrees of freedom during a slow quench of the lattice depth. To this end we compare the temperature $T_s$ extracted from the measured staggered structure factor at half filling to the temperature $T_d$ obtained from fitting NLCE data \cite{Khatami2011,Cocchi2016} to the singles density profiles. We note that $T_s$ measures the local temperature of the gas around half filling whereas $T_d$ is a global measure of the temperature since it is extracted from the entire profile of the cloud. We expect $T_s$ and $T_d$ to agree as long as the spin and density degrees of freedom are globally in thermal equilibrium. Fig.~4 shows the measured temperatures $T_s$ and $T_d$ as a function of the duration $t_\mathrm{quench}$ within which the in-plane lattice depth is increased from $0{E_\mathrm{rec}}$ to $6{E_\mathrm{rec}}$ using a sine-squared ramp shape. For our fastest quench we observe a distinct mismatch between the two temperatures, which gradually diminishes and vanishes for quench times larger than $\unit[0.5]{s}$. We attribute this behavior of $T_d$  to residual compression of the cloud during the quench of the lattice depth due to a slight change of the trapping potential and the compressibility of the gas (see insets of Fig.~4).
The almost linear increase in the spin temperature $T_s$ suggests that the gas reaches \textit{local} equilibrium much earlier. 
This is further supported by the fact that the spin temperature $k_BT_s$ rises with a slope of $0.18(5)\,t/\mathrm{s}$, which is a factor of two smaller than the heating rate $0.37(5)\,t/\mathrm{s}$ that was observed when the thermalized gas was held at the final lattice depth for variable periods of time.

\begin{figure}
	\centering
 \includegraphics[width=0.85\columnwidth,clip=true]{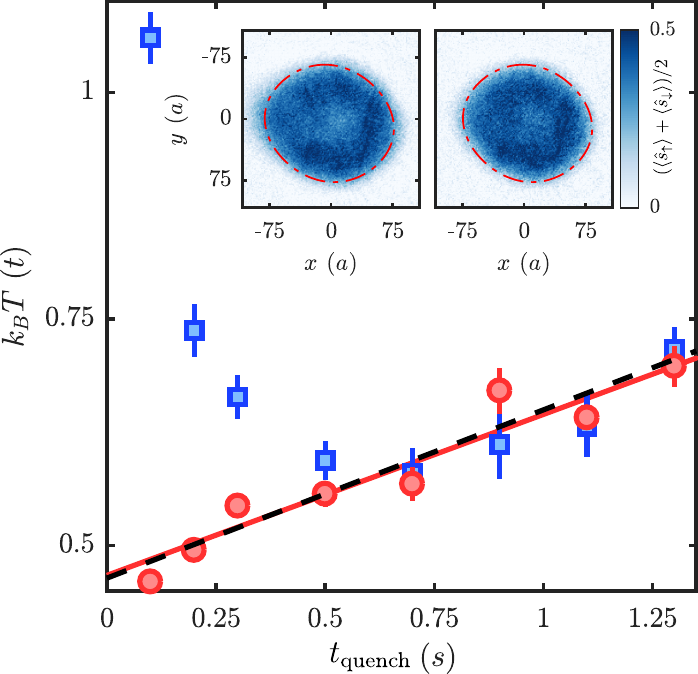}
 \caption{Thermalization of spin and density degrees of freedom. Shown are the density temperature $T_d$  (blue squares) and spin temperature $T_s$ (red circles) as a function of quench time $t_\mathrm{quench}$. Density and spin degrees of freedom reach thermal equilibrium for $t_{\mathrm{quench}} > \unit[0.5]{s}$. The red line is a linear fit to $T_s$ for $t_{\mathrm{quench}} >\unit[0.5]{s}$, which we compare to an extrapolation using half the heating rate measured by holding at the final lattice depth (black dashed line). The insets showing the averaged density profiles at $t_{\mathrm{quench}} = \unit[0.1]{s}$ (left) and $t_{\mathrm{quench}} = \unit[0.5]{s}$ (right) suggest that thermalization is related to density redistribution towards the trap center. The red dash-dotted line is a guide to the eye.}
\label{fig4}
\end{figure}
In conclusion, we have measured the magnetic structure factor of the 2D Hubbard model both momentum- and filling-resolved using a technique based on the coherent manipulation of spin correlations. We have shown that even with moderate imaging resolution, the staggered structure factor, which is ideally suited for thermometry, can be resolved. The technique of imprinting spin patterns of well defined wave vector offers novel possibilities, e.g. {to investigate spin diffusion \cite{Sommer2011,Koschorreck2013,luciuk2017observation}  or to measure the dispersion relation of spin wave excitations in the Hubbard model}.

This work has been supported by BCGS, the Alexander-von-Humboldt Stiftung, ERC (grant 616082), DFG (SFB/TR 185 project B4) and Stiftung der deutschen Wirtschaft.

%\bibliography{thesis}

%

\end{document}